\documentclass[titlepage,12pt]{utarticle}
\usepackage{amsmath,amssymb,amsthm,amscd,cite}
\usepackage{latexsym,amsmath,amssymb,graphicx}
\usepackage[all]{xy}
\usepackage{epsf}
\xyoption{v2}
\xyoption{2cell}
\xyoption{dvips}
\CompileMatrices
\LaTeXdiagrams
\UseAllTwocells
\newdimen\tableauside\tableauside=1.0ex
\newdimen\tableaurule\tableaurule=0.4pt
\newdimen\tableaustep
\def\phantomhrule#1{\hbox{\vbox to0pt{\hrule height\tableaurule width#1\vss}}}
\def\phantomvrule#1{\vbox{\hbox to0pt{\vrule width\tableaurule height#1\hss}}}
\def\sqr{\vbox{%
  \phantomhrule\tableaustep
  \hbox{\phantomvrule\tableaustep\kern\tableaustep\phantomvrule\tableaustep}%
  \hbox{\vbox{\phantomhrule\tableauside}\kern-\tableaurule}}}
\def\squares#1{\hbox{\count0=#1\noindent\loop\sqr
  \advance\count0 by-1 \ifnum\count0>0\repeat}}
\def\tableau#1{\vcenter{\offinterlineskip
  \tableaustep=\tableauside\advance\tableaustep by-\tableaurule
  \kern\normallineskip\hbox
    {\kern\normallineskip\vbox
      {\gettableau#1 0 }%
     \kern\normallineskip\kern\tableaurule}%
  \kern\normallineskip\kern\tableaurule}}
\def\gettableau#1 {\ifnum#1=0\let\next=\null\else
  \squares{#1}\let\next=\gettableau\fi\next}
\numberwithin{equation}{section}
\def\fd{\mathsf{D}}
\def\fe{\mathsf{E}}
\def\fa{\mathsf{A}}

\newcommand{\be}{\begin{equation}}
\newcommand{\ee}{\end{equation}}

\newcommand{\IP}{\mathbb{P}}

\newcommand\IZ{\mathbb {Z}}

\newcommand{\ba}{\begin{array}}
\newcommand{\ea}{\end{array}}

\newcommand{\om}{\overline{M}}

\newcommand{\IF}{{\mathbb F}}

\begin{document}
\preprint{
    {\tt hep-th/0507058}\\
}
\title{
The Ruled Vertex and $\fd$-$\fe$ Degenerations}
\author{Duiliu-Emanuel Diaconescu\footnote{{\tt duiliu@physics.rutgers.edu}}~~and 
Bogdan Florea\footnote{{\tt florea@physics.rutgers.edu}}}
\oneaddress{
      \smallskip
      {\centerline {\it  Department of Physics and Astronomy, 
Rutgers University,}}
      \smallskip
      {\centerline {\it Piscataway, NJ 08854-8019, USA}}}

\date{July 2005}

\Abstract{
We compute all genus topological amplitudes on configurations of ruled surfaces
obtained by resolving lines of $\fd$-$\fe$ singularities in compact Calabi-Yau 
threefolds. We find that our results are in agreement with genus zero 
mirror symmetry calculations, which is further evidence for the validity of the ruled 
vertex formalism for degenerate torus actions. 
}
\maketitle 

\section{Introduction}

A vertex formalism for topological amplitudes in the presence of degenerate 
torus actions has been recently developed in \cite{DFS}. In the present 
note we develop this formalism for configurations of ruled surfaces 
obtained by resolving genus zero curves of $\fd$-$\fe$ singularities 
in compact Calabi-Yau threefolds. These are nontoric exceptional divisors 
in the ambient threefold whose higher genus contribution to the partition function 
could not be determined directly in the ${\bf A}$-model with the methods available so far. 
The computations will be performed for a specific elliptically and $K3$ fibered threefold 
with a line of $\fd_4$ singularities, which is the resolution of the degree $36$ hypersurface 
in $\IP^4_{[1,1,4,12,18]}$. 

The formalism of \cite{DFS} computes the topological partition 
function of local ruled surfaces defined by localization with respect to a 
torus action. More precisely, in order to deal with noncompactness issues 
for the moduli spaces of stable maps, we have to employ localization in order to 
define the theory, not just as a computational tool. 
In more technical terms, we are dealing with residual 
Gromov-Witten theory \cite{local} of a ruled surface embedded in a Calabi-Yau 
threefold. 

Although this theory is mathematically well defined, it is not a priori clear if the 
resulting vertex formalism has concrete physical applications. In this paper 
we will address this question in the context of II{\bf A} compactification on Calabi-Yau 
threefolds with a line of $\fd$-$\fe$ singularities. We will show that the construction of 
\cite{DFS} yields exact results for topological string amplitudes associated to curve 
classes supported on the exceptional locus of the resolution. This is a subsector of 
the full topological partition function of the compact Calabi-Yau threefold. 

We should also mention the implications of our results 
for heterotic - II{\bf A} duality. We obtain 
exact results for topological string amplitudes with nontrivial degree along the 
base of the type II{\bf A} $K3$ fibration in the limit in which the II{\bf A} $K3$ 
is very large keeping the size of exceptional curves finite. 
Interpreted in the heterotic theory, our formulae represent 
exact nonperturbative corrections to the vector multiplet moduli space in a certain 
limit of the heterotic theory which will be described in section 4. 

{\it Acknowledgements}. We would like to thank Eleonora Dell'Aquila for collaboration at an early stage 
of the project and Natalia Saulina for comments on the manuscript. D.-E. D. was partially supported
by an Alfred P. Sloan fellowship and the work of B. F. was partially supported by DOE grant DE-FG02-96ER40949.

\section{$\fd$-$\fe$ Degenerations of Calabi-Yau Threefolds} 

Throughout this paper we will be mainly interested in exceptional configurations of 
ruled surfaces obtained by resolving genus zero curves of 
$\fd_n$, $n\geq 4$ and respectively $\fe_n$, $n=6,7,8$ singularities in Calabi-Yau threefolds. 

Although it is not hard to construct compact Calabi-Yau threefolds with such singularities, 
in this section we will construct a local model which captures the essential features 
of the resolution. It will become shortly clear that the topological partition function 
for curve classes supported on the resolution does not depend on the embedding of 
the exceptional locus in a compact threefold. Therefore we will obtain a 
universal contribution to the topological amplitudes associated to a particular type 
of degeneration. 

The local models will be realized as hypersurfaces ${\widehat X}$ in noncompact toric varieties $Z$ 
with toric presentation of the form 
\[
\begin{array}{ccccc}
Z_1 & Z_2 & U & V & W \cr
1 & 1 & a & b & c \cr
\end{array} 
\]
with disallowed locus $\{Z_1=Z_2=0\}$.
Note that $Z$ is isomorphic to the total space of the rank three bundle $\CO(a)\oplus \CO(b)\oplus \CO(c)$ 
over $\IP^1$. 
We record below the defining equations and the values of $a,b,c$ for singularities in the $\fd$ and $\fe$ 
series
\be\label{eq:DEdeg}
\begin{aligned} 
& \fd_n, \ n\geq 4:& & a= -2(n-1) & & b=-2(n-1) & & c=-4 &  &\quad U^2 + V^2 W = W^{n-1} &\cr
& \fe_6: & & a=-12 & & b=-8& & c=-6& &\quad U^2+V^3+W^4 =0 & \cr
& \fe_7: & & a=-18 & & b=-12 & & c=-8 & & \quad U^2 +V^3 +VW^3 =0 \cr
& \fe_8:& & a = -30 & & b=-20 & & c=-12 & & \quad U^2 +V^3 +W^5 =0. \cr
\end{aligned}
\ee
All hypersurfaces described in \eqref{eq:DEdeg} are singular along the rational curve 
\be\label{eq:sing}
U=V=W=0.
\ee
The resolution $X$ of a singular hypersurface ${\widehat X}$ in the above series can be obtained by
performing toric blow-ups of the ambient variety $Z$. One can easily check that $X$ has the structure 
of a fibration in affine surfaces over $\IP^1$ with typical fiber isomorphic to the canonical resolution of
a $\fd$-$\fe$ surface singularity. Since the singularity type does not jump along the curve \eqref{eq:sing}, 
there will not be any special fibers. Therefore each fiber contains a chain of rational curves 
$f_0,\ldots, f_{n-1}$ which intersect according to the corresponding Dynkin diagram. 
Taking into account the fibration structure over $\IP^1$, the curves $f_0,\ldots, f_{n-1}$ will 
generate a collection of ruled surfaces $S_0,\ldots, S_{n-1}$ in $X$ in one to one 
correspondence with the nodes of the Dynkin diagram. These surfaces intersect along common sections 
as shown in the figure below.

\begin{figure}[ht]
\centerline{\epsfxsize=9.8cm \epsfbox{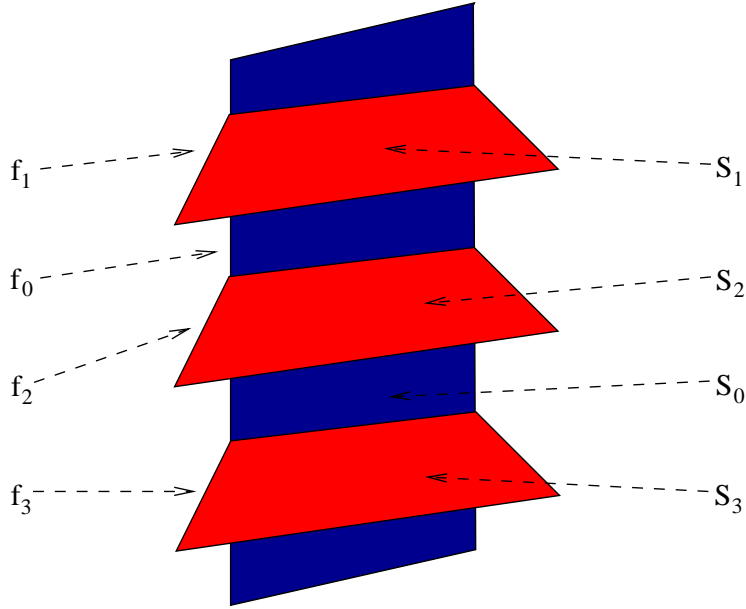}}
\caption{$\fd_4$ configuration of ruled surfaces.} 
\label{fig:D4} 
\end{figure}

We will fix the notation so that $S_1$ denotes the central component of the exceptional locus 
which meets three other components along disjoint sections. Since $S_0$ is an irreducible ruling, 
it follows that it must be isomorphic to $\IF_0=\IP^1\times \IP^1$. Otherwise $S_0$ would not have 
three disjoint sections. The degree of the remaining surfaces can be determined inductively starting 
from the central component. If two surfaces $S_i, S_j$ meet along a common section $C_{ij}$, 
according to the adjunction formula we have 

\[ 
\hbox{deg}(N_{S_i}C_{ij}) + \hbox{deg}(N_{S_j}C_{ij}) = -2. 
\]
Using successively this formula we can determine the degrees $e_i$ of the ruled surfaces $S_i$ as shown 
in fig. 1. 

It is important to note for our purposes that the above configuration of ruled surfaces is not toric, but 
it admits a degenerate torus action\footnote{Adopting the terminology introduced in \cite{DFS}, 
we will call a torus action on a Calabi-Yau threefold nondegenerate if the fixed locus 
is zero dimensional, and degenerate if the zero locus is higher dimensional.}. The configuration is not toric 
because the central component 
$S_0 \simeq \IF_0$ contains three marked sections $C_{01}, C_{02}, C_{03}$ corresponding to the intersection 
loci with $S_1, S_2, S_3$. Therefore any torus action on $X$ preserving the exceptional divisor 
$S_0+\ldots + S_{n-1}$ must act with weight zero on the fiber $f_0$ of the central component $S_0$.  

Let us consider a torus action with weights $\lambda_{f_i}$, $i=0,\ldots,n-1$ 
on the fibers $f_0,\ldots,f_{n-1}$ of the rulings and $\lambda_{C_{ij}}$ on the sections $C_{ij}$. 
As noted above, $\lambda_{f_0}=0$. In the following we will consider only torus actions satisfying the 
extra condition 
\[ 
\lambda_{C_{0i}} + \lambda_{f_i} =0, \qquad i=1,2,3.
\]
This means that the fixed fibers $f_0,f_0'$ are locally equivariant Calabi-Yau in the terminology 
of \cite{local}. 

Given such a torus action one can employ the vertex formalism developed in \cite{DFS} 
in order to compute the topological partition function. A priori this is a generating functional 
for residual Gromov-Witten invariants, which are defined \cite{local} by summing over 
fixed loci in the moduli space of stable maps to $X$. In particular the invariants will depend
on the choice of a torus action. However, in the present situation, one can check that the 
moduli space of stable maps $\om_{g,0}(X,\beta)$ is compact if $\beta$ is a curve class supported on 
the the exceptional locus. Therefore the result has to be independent of the torus action, and moreover
it has to agree with the Gromov-Witten invariants of any compact Calabi-Yau threefold which contains 
a degeneration of this form. We will check this statement by explicit computations in section 4. 

Before explaining the construction of the topological partition function, let us briefly discuss 
the case of $\fa_n$ degenerations. One might wonder why we chose to exclude this case from our considerations 
given the fact that these are the simplest degenerations from a geometric point of view. 
If we do not allow the singularity type to jump, an $\fa_n$ degeneration will give rise to a toric 
configuration of ruled surfaces, and the topological partition function can be computed using 
the topological vertex formalism \cite{topvert}. If we do allow the singularity type to jump, 
the resolution will contain chains of ruled surfaces with reducible fibers which were considered before 
in \cite{DFS}. Typically, in such cases the residual Gromov-Witten invariants will depend on the 
choice of a torus action, therefore we would not obtain a universal contribution to the partition function 
of a generic Calabi-Yau threefold containing the degeneration in question.
In conclusion the $\fd$ and $\fe$ series introduced above seem to be the most interesting testing ground 
for ideas discussed in this paper. 

\section{The Topological Partition Function for $\fd$-$\fe$ Degenerations} 

In this section we explain the construction of the topological partition function for $\fd$ and $\fe$ degenerations. 
Since the central idea is the same in all examples it suffices to present one model in detail. 

Let us start with the $\fd_4$ degeneration
We have a central component $S_0\simeq \IF_0$ and three additional components $S_1,S_2,S_3 \simeq \IF_2$. 
$S_0$ intersects each of the other components along a section $C_{0i}$, $i=1,2,3$. Note that $C_{0i}$ is the 
negative section on each $S_i$, $i=1,2,3$ which squares to $(-2)$. Taken in isolation, the partition function 
of each component can be easily written in the topological vertex formalism \cite{topvert}. 
For the central component $S_0$ we have 
\be\label{eq:Fzero} 
Z_{S_0} = \sum_{R,R',Q,Q'} W_{RQ}W_{QR'}W_{R'Q'}W_{Q'R} q_{f_0}^{l(R)+l(R')} q_b^{l(Q)+l(Q')}
\ee
where $l(Q)$ is the total number of boxes of the Young diagram $Q$ 
and $W_{RQ}$ are functions of $q=e^{ig_s}$ defined 
as the large $N$ limit of the S-matrix 
of Chern-Simons theory
\[ 
W_{RQ}(q) = \hbox{lim}_{N\to \infty} q^{-{N(l(R)+l(Q))\over 2}} 
{S_{RQ}(q,N)\over S_{00}(q,N)}.
\]
Note that $W_{PR}$ is symmetric in $(P,R)$ and
can be written  in terms of Schur functions $s_R$:
\[
 W_{P R}(q)=s_{R}\left(q^{-i+1/2} \right)
s_{P}\left(q^{R^i-i+1/2} \right) 
\]
where $R^i$ is the length of the i-th row of Young diagram $R$
and $i=1,\ldots, \infty.$ The variables $q_{f_0}, q_b$ are the exponentiated K\"ahler parameters associated 
to the fiber class $f_0$ and respectively the section class $[C_{0i}]$, $i=1,2,3$. 
The expression of the partition function for a local $\IF_2$ surface is similar, except for 
some additional framing factors 
\be\label{eq:Ftwo}
Z_{S_i} = \sum_{S,S',P,P'} q^{-\kappa(P)}q^{\kappa(P')}  
W_{SP}W_{PS'}W_{P'Q'}W_{Q'P} q_{f_i}^{l(S)+l(S')+2l(P')} q_b^{l(P)+l(P')}
\ee
where $\kappa(R)$ is defined by the formula 
\be\label{eq:kappa} 
\kappa(R) = 2\sum_{\tableau{1}\in R} (i(\tableau{1})-j(\tableau{1}))
\ee
in which $i(\tableau{1}), j(\tableau{1})$ specify the position of a given box 
in the Young diagram. 
The variables $q_{f_i}$, $i=1,2,3$ are 
the exponentiated K\"ahler parameter associated to the fiber classes $f_i$, $i=1,2,3$. 

In order to write down the partition function for the collection of surfaces $S_0+\ldots+ S_4$, we have to glue
together the building blocks \eqref{eq:Fzero} and \eqref{eq:Ftwo}. At this point we have to use the ruled 
surface vertex found in \cite{DFS} since the divisor $S_0+\ldots + S_4$ is not toric. 
Let us briefly recall the construction of \cite{DFS}.  

The topological partition function for a 
local ruled surface was obtained by a TQFT algorithm based on a decomposition of the ruled surface in 
basic building blocks. Given a ruled surface $S$ over a curve $\Sigma$ 
of genus $g$ (possibly with reducible fibers), 
the decomposition in question is induced by a decomposition of the base $\Sigma$ in pairs of 
pants and caps. The caps may be absent if $S$ is an irreducible ruling, but they are required 
if $S$ has reducible fibers. We will not need the caps in the following.  

To a pair of pants $\Delta$ occurring in the decomposition of $\Sigma$ we associate a piece of a 
ruled surface obtained be restricting the fibration $S\to \Sigma$ to $\Delta$. Therefore we obtain a
$\IP^1$ bundle $P$ over $\Delta$ which is is characterized by an integer level $p$. 
The case we will need below corresponds to $p=0$. Note that there is a degenerate torus action 
on $P$ fixing to sections of the ruling. The topological partition function associated to the 
level zero ruled 
vertex $P$ is a formal expression $V_{RR'}$
which depends on two arbitrary Young diagrams $R,R'$ associated to the 
fixed sections. According to \cite{DFS},  $V_{RR'}$ is given by the following expression 
\be\label{eq:ruledvertex} 
V_{RR'} = \left(\sum_Q W_{RQ} W_{QR'} q_r^{l(Q)}\right)^{-1} 
\ee
where $q_r$ is the K\"ahler parameter of the ruling. 

We now return to the construction of the partition function of the $\fd_4$ degeneration. 
Note that if we excise the three marked sections $C_{0i}$, $i=1,2,3$ from the central component 
$\IF_0$, we obtain precisely a ruled vertex $P=\Delta \times \IP^1$. Therefore the partition function will 
be obtained by gluing the contributions of the three components $S_i$, $i=1,2,3$ to 
the ruled vertex \eqref{eq:ruledvertex}. An informal way of describing the gluing is by virtually decomposing
the exceptional locus into basic building blocks using noncompact branes as shown in fig. 2. 

\begin{figure}[ht]
\centerline{\epsfxsize=16.1cm \epsfbox{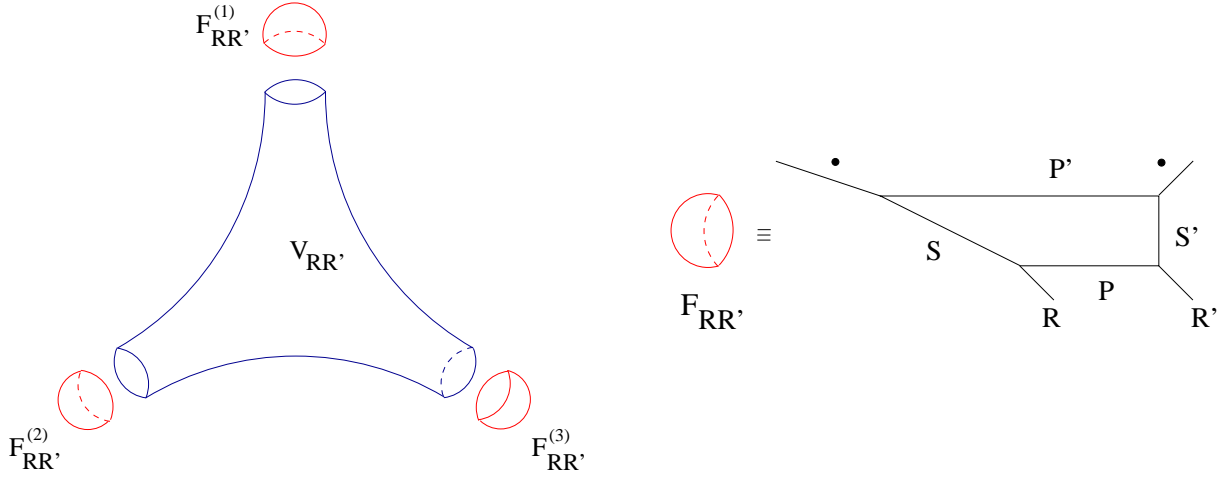}}
\caption{$\fd_4$ lego. The pair of pants represents the ruled vertex defined in \cite{DFS}; the caps 
represent the contribution of the $\IF_2$ surfaces.} 
\label{fig:D4lego} 
\end{figure}

This makes it clear that in the process of gluing we will have to use a topological vertex with all 
three legs nontrivial for each point of intersection between the sections $C_{0i}$ and the sections 
$f_0, f_0'$ fixed by the torus action. Collecting all the pieces we obtain the following expression 
\be\label{eq:DfourA} 
Z_{\fd_4} = \sum_{R,R'} V_{R^tR'} q_{f_0}^{l(R)+l(R')}\prod_{i=1}^3 F_{RR'}(q_{f_i})
\ee
where $F(q_i)$ is the following formal expression 
\be\label{eq:DfourB} 
\begin{aligned} 
F_{RR'}(q_{f_i}) = \sum_{S,S',P,P'} & C_{RSP^t} C_{S^t\bullet (P')^t}C_{P'\bullet(S')^t} 
C_{S'R'P}q^{-\kappa(R)/2}q^{\kappa(S)/2}q^{-\kappa(S')/2}
q^{-\kappa(P)/2}q^{\kappa(P')/2} \cr
& \times q_{f_i}^{l(S)+l(S')+2l(P')} q_b^{l(P)+l(P')}.\cr
\end{aligned} 
\ee

The partition function for other $\fd_n$ as well as $\fe_n$ degenerations can be easily written by analogy. 
The contribution of the central component $S_0\simeq \IF_0$ is the same in all cases. Therefore the 
generic form of the partition function is 
\[ 
\sum_{R,R'} V_{R^tR'} q_{f_0}^{l(R)+l(R')}\prod_{i=1}^3 F_{RR'}^{(i)}(q_{f_i}) 
\] 
where the functions $F_{RR'}^{(i)}(q_{f_i})$ represent the contributions of the remaining three 
chains of ruled surfaces. Since these chains are toric, $F_{RR'}^{(i)}(q_{f_i})$ can be obtained 
using the topological vertex formalism as in \cite{IKPi}. 

For further reference, we will write down the first few terms in the expansion of the topological free 
energy for the $\fd_4$ singularity.  Although the formulas \eqref{eq:DfourA}, 
\eqref{eq:DfourB} are valid for arbitrary K\"ahler parameters, for the purpose of comparison with 
mirror symmetry calculations in a compact Calabi-Yau model, it suffices to set  
$q_{f_1}=q_{f_2}=q_{f_3}=q_f$. 
For notational simplicity we will also set $q_{f_0}=q_0$. 
Then up to degree two in $q_b,q_0$ and degree three in $q_f$, the free energy expansion reads 
\be\label{eq:DfourC} 
\begin{aligned} 
F_{\fd_4} = 
& -{1\over \left(2\hbox{sin}{g_s\over 2}\right)^2}(2q_b+2q_0+6q_f+4q_bq_0+6q_bq_f+6q_0q_f) \cr
& -{1\over \left(2\hbox{sin}{g_s\over 2}\right)^2}(6 q_b^2q_0+6q_bq_0^2 + 18 q_bq_0q_f + 
12 q_bq_f^2 + 6 q_0 q_f^2) \cr
& -{1\over \left(2\hbox{sin}{g_s\over 2}\right)^2}(32 q_b^2q_0^2 +
30 q_b^2q_0q_f + 30 q_bq_0^2q_f + 42 q_bq_0q_f^2 + 
18 q_bq_f^3 + 2q_0q_f^3) + 9 q_b^2q_0^2\cr
& -{1\over \left(2\hbox{sin}{g_s\over 2}\right)^2}(210q_b^2q_0^2q_f + 84 q_b^2q_0q_f^2+84 q_bq_0^2q_f^2
+18 q_b^2q_f^3 + 76 q_bq_0q_f^3 + 2 q_0^2q_f^3) + 48 q_b^2q_0^2q_f\cr\nonumber
\end{aligned}
\ee
\be\label{eq:DfourCi}
\begin{aligned}
& -{1\over \left(2\hbox{sin}{g_s\over 2}\right)^2}(720q_b^2q_0^2q_f^2 + 222 q_b^2q_0q_f^3 + 174 q_bq_0^2q_f^3)
+147 q_b^2q_0^2q_f^2\cr
& -{1850 q_b^2q_0^2q_f^3 \over \left(2\hbox{sin}{g_s\over 2}\right)^2}+ 336 q_b^2q_0^2q_f^3 + \
\hbox{multicover contributions}\ + \cdots~. \cr
\end{aligned} 
\ee
One can also check that the multicover contributions have the expected BPS behavior. 
In the next section we will compare the genus zero terms in this expressions with the 
prepotential of an elliptically fibered compact Calabi-Yau model which 
contains generically a line of $\fd_4$ singularities. 

\section{A Compact Calabi-Yau Model with $\fd_4$ Singularities} 

In order to test the ruled vertex formalism, we would like next to compare the expansion 
\eqref{eq:DfourC} with the topological free energy of a compact Calabi-Yau model which exhibits a 
$\fd_4$ degeneration as in fig. 1. The most familiar examples are the elliptic fibrations over Hirzebruch 
surfaces $\IF_n$ with lines of $\fd$ or $\fe$ singularities. As explained in \cite{BIKMSV}, a generic 
$\fd_4$ singularity will correspond to $G_2$ enhanced symmetry with matter. To obtain the $SO(8)$ 
configuration of surfaces, we have to take $n=4$, which will correspond to an $SO(8)$ theory 
without matter. 

The vertices of the $\nabla$ polytope are as follows:
$$
{\tt V_1=(-1,0,2,3),~V_2=(0,0,-1,0),~V_3=(0,0,0,-1),~V_4=(1,4,2,3),~V_5=(0,-1,2,3).}
$$
\noindent The Calabi-Yau hypersurface $X$ has $h^{1,1}(X)=7,~h^{2,1}(X)=271$\footnote{We have used the program 
{\tt POLYHEDRON} written by Philip Candelas.}, in agreement with the 
existence of an $SO(8)$ in the Picard lattice. $X$ can also be viewed as the resolution of the degree $36$ 
hypersurface in $\IP^4_{[1,1,4,12,18]}$. The other toric Calabi-Yau divisors are associated 
with the following points in the $\nabla$ polytope:
$$
{\tt B=(0,0,2,3),~D_0=(0,2,2,3),~D_1=(0,1,1,1),~D_{\infty}=(0,1,2,3)}.
$$
The Mori cone of the Calabi-Yau $X$ is obtained employing the method of the piecewise linear 
functions \cite{OP}:
\be\label{eq:Mcone}
\begin{array}{cccccccccc}
& V_1 & V_2 & V_3 & V_4 & V_5 & B & D_0 & D_1 & D_{\infty} \cr
e^{(1)}: & 0 & 1 & 0 & 0 & 0 & 0 & -2 & 3 & 1 \cr
e^{(2)}: & 0 & 0 & 1 & 0 & -2 & 3 & 0 & -2 & 0  \cr
e^{(3)}: & 0 & 0 & 0 & 0 & 0 & 1 & 1 & 0 & -2  \cr
e^{(4)}: & 0 & 0 & 0 & 0 & 1 & -2 & 0 & 0 & 1 \cr
e^{(5)}: & 1 & 0 & 0 & 1 & 0 & 0 & -2 & 0 & 0 \cr
\end{array}
\ee
and is five dimensional; the same is true for its dual K\"ahler cone. The K\"ahler cone of the ambient toric variety 
$Z$ is nine dimensional. If we denote by $\iota: X\hookrightarrow Z$ the inclusion map, we have the following 
exact sequence
$$
0\rightarrow {\cal K}\rightarrow H^2(Z)\stackrel{\iota^*}{\rightarrow} H^2(X)\rightarrow {\cal C}\rightarrow 0.
$$
\noindent The image of the map $\iota^*: H^2(Z)\rightarrow H^2(X)$ is five dimensional, and it has a four dimensional 
kernel ${\cal K}$ and two dimensional cokernel ${\cal C}$. Therefore, all of the exceptional surfaces $S_1,S_2$ and 
$S_3$ stay in the same divisor class in the toric Calabi-Yau hypersurface, which we identify with $D_1$. We also identify 
$D_0$ with the divisor class of the central component, $D_{\infty}$ with the divisor 
class of the affine component and $B$ with the section of the elliptic fibration.  

Next, we perform a mirror symmetry computation to check the genus zero invariants obtained in the previous 
section. In order to do this, we need to identify the curve classes we are interested in with positive integer 
linear combinations of the generators of the Mori cone. We arrive at the following relations:
\be
\begin{aligned}
&f_0=e^{(1)},~~f_1=f_2=f_3=f=e^{(2)}+e^{(3)}+2e^{(4)},~~f_{\infty}=e^{(3)},~~
C_0=e^{(5)},\cr
&C_0^{\IF_4}=2e^{(3)}+e^{(5)},~~f^{\IF_4}=e^{(4)}, \nonumber
\end{aligned}
\ee
where $C_0^{\IF_4}$ and $f^{\IF_4}$ are the negative section and fiber classes of the base $\IF_4$ 
respectively. We note that the exceptional fiber class fractionates in the ambient toric variety, and therefore, 
in order to compute the genus zero invariants in \eqref{eq:DfourC} we need to compute the prepotential 
up to order $16$ in the instanton expansion. The generators of the K\"ahler cone are given by
\be
\begin{aligned}
& {\cal R}_1=12V_4+2B+6D_0+3D_1+4D_{\infty},~~{\cal R}_2=18V_4+3B+9D_0+4D_1+6D_{\infty},~~{\cal R}_5=V_4,\cr
& {\cal R}_3=26V_4+4B+13D_0+6D_1+8D_{\infty},~~{\cal R}_4=40V_4+6B+20D_0+9D_1+13D_{\infty},\nonumber
\end{aligned}
\ee
and the K\"ahler form is $J=\sum_{i=1}^5t_i{\cal R}_i$. The nontrivial triple intersections numbers are \cite{KS}
\be
\begin{aligned}
& D_0^3=8, & & D_0^2D_{\infty}=-2, & & D_{\infty}^3=8, & & D_0^2D_1=-6, & & D_1^3=24, &\cr
& BD_{\infty}^2=4,& &B^2D_{\infty}=2,& &B^3=8,& &D_0^2V_4=-2,& &D_0D_{\infty}V_4=1,& \cr
& D_{\infty}^2V_4=-2,& &D_0D_1V_4=3,& &D_1^2V_4=-6,& &BD_{\infty}V_4=1,& &B^2V_4=-2.&
\end{aligned}
\ee

The fundamental period for the mirror of $X$ is given by
\be
\begin{aligned}
\varpi_0&=\sum_{n_1,n_2,n_3,n_4,n_5}h_0\cr
&=\sum_{n_1,n_2,n_3,n_4,n_5}\frac{z_1^{n_1}z_2^{n_2}z_3^{n_3}z_4^{n_4}z_5^{n_5}}{\Gamma(1+n_1)\Gamma(1+n_2)\Gamma(1+n_5)^2
\Gamma(1-2n_2+n_4)\Gamma(1+3n_2+n_3-2n_4)}\cr
&\hspace{2.1cm}\times\frac{\Gamma(1+3n_1)}{\Gamma(1-2n_1+n_3-2n_5)\Gamma(1+3n_1-2n_2)\Gamma(1+n_1-2n_3+n_4)},
\end{aligned}
\ee
where $z_i$, $i=1,\ldots,5$ are the large complex structure coordinates. We denote by $\varpi_i$, $i=1,\ldots,5$ the 
periods that behave asymptotically like $\ln z_i$, $i=1,\ldots,5$; these are given by
\be
\varpi_i=\varpi_0\ln z_i+g_i,~~i=1,\ldots,5,
\ee
where 
\be
\begin{aligned}
g_1&=\sum_{n_1,n_2,n_3,n_4,n_5}[h_0(3S_{3n_1}-S_{n_1}+2S_{-2n_1+n_3-2n_5}-3S_{3n_1-2n_2}-S_{n_1-2n_3+n_4})+2h_3-3h_4-h_5],
\nonumber
\end{aligned}
\ee
\be
\begin{aligned}
g_2&=\sum_{n_1,n_2,n_3,n_4,n_5}[h_0(-S_{n_2}+2S_{-2n_2+n_4}-3S_{3n_2+n_3-2n_4}+2S_{3n_1-2n_2})+2h_1-3h_2+2h_4],\cr
g_3&=\sum_{n_1,n_2,n_3,n_4,n_5}[h_0(-S_{3n_2+n_3-2n_4}-S_{-2n_1+n_3-2n_5}+2S_{n_1-2n_3+n_4})-h_2-h_3+2h_5],\cr
g_4&=\sum_{n_1,n_2,n_3,n_4,n_5}[h_0(-S_{n_4}+2S_{3n_2+n_3-2n_4}-S_{n_1-2n_3+n_4})-h_1+2h_2-h_5],\cr
g_5&=\sum_{n_1,n_2,n_3,n_4,n_5}[h_0(-2S_{n_5}+2S_{-2n_1+n_3-2n_5})+2h_3].
\end{aligned}
\ee
In the above we have introduced the notation $S_n=\sum_{k=1}^n\frac{1}{k}$ and we have defined
\be
\begin{aligned}
h_1=&\frac{z_1^{n_1}z_2^{n_2}z_3^{n_3}(-z_4)^{n_4}z_5^{n_5}}{\Gamma(1+n_1)\Gamma(1+n_2)\Gamma(1+n_5)^2
\Gamma(1+3n_2+n_3-2n_4)\Gamma(1-2n_1+n_3-2n_5)}\cr
&\times\frac{\Gamma(1+3n_1)\Gamma(2n_2-n_4)}{\Gamma(1+3n_1-2n_2)\Gamma(1+n_1-2n_3+n_4)},\cr
h_2=&\frac{z_1^{n_1}(-z_2)^{n_2}(-z_3)^{n_3}z_4^{n_4}z_5^{n_5}}{\Gamma(1+n_1)\Gamma(1+n_2)\Gamma(1+n_5)^2
\Gamma(1-2n_2+n_4)\Gamma(1-2n_1+n_3-2n_5)}\cr
&\times\frac{\Gamma(1+3n_1)\Gamma(-3n_2-n_3+2n_4)}{\Gamma(1+3n_1-2n_2)\Gamma(1+n_1-2n_3+n_4)},\cr
h_3=&\frac{z_1^{n_1}z_2^{n_2}(-z_3)^{n_3}z_4^{n_4}z_5^{n_5}}{\Gamma(1+n_1)\Gamma(1+n_2)\Gamma(1+n_5)^2
\Gamma(1-2n_2+n_4)\Gamma(1+3n_2+n_3-2n_4)}\cr
&\times\frac{\Gamma(1+3n_1)\Gamma(2n_1-n_3+2n_5)}{\Gamma(1+3n_1-2n_2)\Gamma(1+n_1-2n_3+n_4)},\cr
h_4=&\frac{(-z_1)^{n_1}z_2^{n_2}z_3^{n_3}z_4^{n_4}z_5^{n_5}}{\Gamma(1+n_1)\Gamma(1+n_2)\Gamma(1+n_5)^2
\Gamma(1-2n_2+n_4)\Gamma(1+3n_2+n_3-2n_4)}\cr
&\times\frac{\Gamma(1+3n_1)\Gamma(-3n_1+2n_2)}{\Gamma(1-2n_1+n_3-2n_5)\Gamma(1+n_1-2n_3+n_4)},\cr
h_5=&\frac{(-z_1)^{n_1}z_2^{n_2}z_3^{n_3}(-z_4)^{n_4}z_5^{n_5}}{\Gamma(1+n_1)\Gamma(1+n_2)\Gamma(1+n_5)^2
\Gamma(1-2n_2+n_4)\Gamma(1+3n_2+n_3-2n_4)}\cr
&\times\frac{\Gamma(1+3n_1)\Gamma(-n_1+2n_3-n_4)}{\Gamma(1-2n_1+n_3-2n_5)\Gamma(1+3n_1-2n_2)}.
\end{aligned}
\ee
Following \cite{CdFKM,HKTY}, we can now write down the mirror map
\be
t_i=-\frac{\varpi_i}{\varpi_0}=-\ln z_i-\frac{g_i}{\varpi_0},~~i=1,\ldots,5.
\ee

In order to compute the genus zero Gromov-Witten invariants of $X$, we need to define the following series
\be
\varpi_{ij}=\left(\frac{\partial^2\Pi}{\partial\rho_i\partial\rho_j}\right)_{|_{\rho_i=\rho_j=0}}~~i,j=1,\ldots,5,
\ee
where
\be
\begin{aligned}
\Pi=&\sum_{n_1,n_2,n_3,n_4,n_5}\prod_{i=1}^5z_i^{n_i+\rho_i}\frac{\Gamma(1+3n_1+3\rho_1)}{\Gamma(1+3\rho_1)}
\frac{\Gamma(1+\rho_1)}{\Gamma(1+n_1+\rho_1)}\frac{\Gamma(1+\rho_2)}{\Gamma(1+n_2+\rho_2)}
\frac{\Gamma(1+\rho_5)^2}{\Gamma(1+n_5+\rho_5^2)}\cr
&\times\frac{\Gamma(1-2\rho_2+\rho_4)}{\Gamma(1-2n_2+n_4-2\rho_2+\rho_4)}\frac{\Gamma(1+3\rho_2+\rho_3-2\rho_4)}
{\Gamma(1+3n_2+n_3-2n_4+3\rho_2+\rho_3-2\rho_4)}\cr
&\times\frac{\Gamma(1-2\rho_1+\rho_3-2\rho_5)}{\Gamma(1-2n_1+n_3-2n_5-2\rho_1+\rho_3-2\rho_5)}
\frac{\Gamma(1+3\rho_1-2\rho_2)}{\Gamma(1+3n_1-2n_2+3\rho_1-2\rho_2)}\cr
&\times\frac{\Gamma(1+\rho_1-2\rho_3+\rho_4)}{\Gamma(1+n_1-2n_3+n_4+\rho_1-2\rho_3+\rho_4)}.
\end{aligned}
\ee
We will not explicitly write down here the series $\varpi_{ij},i,j=1,\ldots,5$. They are available upon request.

Taking into account the triple intersection numbers $(4.2)$, a consequence of mirror symmetry is that the following equations 
hold true (see e.g. \cite{HKTY}):
\be
\begin{aligned}
\partial_{t_1}{\cal F}_X^{(0)}=&-36\varpi_{11}-108\varpi_{12}-156\varpi_{13}-240\varpi_{14}-6\varpi_{15}
-81\varpi_{22}-234\varpi_{23}-360\varpi_{24}\cr
&-9\varpi_{25}-169\varpi_{33}-520\varpi_{34}-13\varpi_{35}-400\varpi_{44}-20\varpi_{45},\cr
\partial_{t_2}{\cal F}_X^{(0)}=&-54\varpi_{11}-162\varpi_{12}-234\varpi_{13}-360\varpi_{14}-9\varpi_{15}
-120\varpi_{22}-348\varpi_{23}-534\varpi_{24}\cr
&-12\varpi_{25}-252\varpi_{33}-774\varpi_{34}-18\varpi_{35}-594\varpi_{44}-27\varpi_{45},\cr
\partial_{t_3}{\cal F}_X^{(0)}=&-78\varpi_{11}-234\varpi_{12}-338\varpi_{13}-520\varpi_{14}-13\varpi_{15}
-174\varpi_{22}-504\varpi_{23}-774\varpi_{24}\cr
&-18\varpi_{25}-364\varpi_{33}-1120\varpi_{34}-26\varpi_{35}-860\varpi_{44}-40\varpi_{45},\cr
\partial_{t_4}{\cal F}_X^{(0)}=&-120\varpi_{11}-360\varpi_{12}-520\varpi_{13}-800\varpi_{14}-20\varpi_{15}
-267\varpi_{22}-774\varpi_{23}-1188\varpi_{24}\cr
&-27\varpi_{25}-560\varpi_{33}-1720\varpi_{34}-40\varpi_{35}-1320\varpi_{44}-60\varpi_{45},\cr
\partial_{t_5}{\cal F}_X^{(0)}=&-3\varpi_{11}-9\varpi_{12}-13\varpi_{13}-20\varpi_{14}
-6\varpi_{22}-18\varpi_{23}-27\varpi_{24}-13\varpi_{33}-40\varpi_{34}\cr
&-30\varpi_{44}.
\end{aligned}
\ee
where ${\cal F}_X^{(0)}$ is the genus zero free energy of $X$. These equations completely determine the prepotential; after a long 
computation we obtain the instanton expansion presented below:
\be
\begin{aligned}
{\cal F}_X^{(0)}=&-2q_1-2q_5-\frac{1}{4}q_1^2-\frac{1}{4}q_5^2-4q_1q_5-6q_1^2q_5-6q_1q_5^2-6q_2q_3q_4^2-\frac{65}{8}q_1^2q_5^2
-6q_2q_3q_4^2(q_1+q_5)\cr
&-18q_1q_2q_3q_4^2q_5-30q_1^2q_2q_3q_4^2q_5-30q_1q_2q_3q_4^2q_5^2-\frac{3}{4}q_2^2q_3^2q_4^4-210q_1^2q_2q_3q_4^2q_5^2-6q_1q_2^2q_3^2q_4^4\cr
&-12q_2^2q_3^2q_4^4q_5-\frac{3}{4}q_2^2q_3^2q_4^4(q_1^2+q_5^2)-42q_1q_2^2q_3^2q_4^4q_5-84q_1^2q_2^2q_3^2q_4^4q_5-84q_1q_2^2q_3^2q_4^4q_5^2\cr
&-\frac{2889}{4}q_1^2q_2^2q_3^2q_4^4q_5^2-\frac{2}{9}q_2^3q_3^3q_4^6-2q_1q_2^3q_3^3q_4^6-18q_2^3q_3^3q_4^6q_5
-2q_1^2q_2^3q_3^3q_4^6-18q_2^3q_3^3q_4^6q_5^2\cr
&-76q_1q_2^3q_3^3q_4^6q_5-174q_1^2q_2^3q_3^3q_4^6q_5-222q_1q_2^3q_3^3q_4^6q_5^2-1850q_1^2q_2^3q_3^3q_4^6q_5^2+\cdots,
\end{aligned}
\ee
where $q_i=e^{-t_i}$, $i=1,\ldots,5$.
Taking into account the relations
\be
q_0=e^{-t_1},~~q_f=e^{-(t_2+t_3+2t_4)},~~q_b=e^{-t_5},
\ee 
we find that the genus zero expansion of the compact model prepotential agrees with the expression \eqref{eq:DfourC}. This is very convincing 
evidence for the ruled vertex formalism strengthening the enumerative tests performed in \cite{DFS}. 

We would like to conclude this section with a brief discussion of the implications of our results for 
heterotic-type II{\bf A} duality. The compact model considered here has the interesting property of 
being related by string duality to two distinct heterotic theories \cite{Kthree}. One possible heterotic dual is the 
$E_8\times E_8$ heterotic string compactified on $K3\times T^2$ with instanton numbers $(8,16)$ and 
unbroken gauge group $SO(8)$ corresponding to the line of $\fd_4$ singularities. Another heterotic dual is 
a $Spin(32)/\IZ_2$ heterotic string on $K3\times T^2$ with instanton number $24$ and unbroken 
$SO(8)$ gauge group. In both cases we have no charged multiplets. 

In section three we have obtained exact results for topological amplitudes supported on the exceptional locus 
of the $\fd_4$ degeneration. This is a particular limit of the theory in which we are sending the size 
of the $K3$ fiber and the elliptic fiber to $\infty$ keeping the sizes of the base and the exceptional divisors 
fixed. Note that this is not the typical weak coupling heterotic limit which has been studied intensively in the 
string duality literature \cite{KV,KLM,AGNT,HM,MM,TK}. In fact it is an easy exercise to understand the corresponding
limit of the heterotic theory using the duality map explained in section 5 of \cite{Kthree}.
For simplicity, we will set the $B$-field on the two-cycles of the II{\bf A} Calabi-Yau space to zero. 
This corresponds to a square heterotic two-torus. Then the limit consists of taking one of the radii 
of the two torus to infinity keeping the other radius, as well as the $SO(8)$ Wilson lines fixed.
We also keep the heterotic dilaton-axion $S=t_b$ fixed. 
In this limit the higher order terms in $q_b$ in formula \eqref{eq:DfourC} can be interpreted as 
nonperturbative corrections to the vector multiplet moduli space depending on the Wilson line moduli. 
Similar results have been obtained before for generic K\"ahler parameters in \cite{KKRS} up to 
genus two on the base. Our formulas include all genus corrections on the base of the II{\bf A} threefold, but 
they are valid only in the limit described above. In would be very interesting to further explore the 
consequences of our formulae for the heterotic string, especially in the strong coupling 
regime.

\bibliography{strings,m-theory,susy,largeN}
\bibliographystyle{utphys}

\end{document}